# Gravitational eigenstates in weak gravity II: further approximate methods for decay rates

**A D Ernest**

Faculty of Science, Charles Sturt University, Wagga Wagga, 2678, Australia

Email: aernest@csu.edu.au

**Abstract**
This paper develops further approximate methods for obtaining the dipole matrix elements and corresponding transition and decay rates of the high-$n$, high-$l$ gravitational eigenstates. These methods include (1) investigation of the polar spreads of the angular components of the high-$n$, high-$l$ eigenstates and the effects these have on the limiting values of the angular components of the dipole matrix elements in the case of large $l$ and $m$ and (2) investigation of the rapid cut off and limited width of the low-$p$, high-$n$ radial eigenfunctions, and the development of an equation to determine the width, position and oscillatory behaviour of those eigenfunctions in cases of arbitrarily large values of $n$, $l$ and $p$. The methods have wider applicability than dipole transition rate estimates and may be also used to determine limits on the rates for more general interactions. Combining the methods enables the establishment of upper limits to the total dipole decay rates of many high-$n$, low-$p$ states on the state diagram to be determined, even those that have many channels available for decay. The results continue to support the hypothetical existence of a specialized set of high-$n$, low-$p$ gravitational eigenfunctions that are invisible and stable, both with respect to electromagnetic decay and gravitational collapse, making them excellent dark matter candidates.

PACS numbers: 03.65.Ge, 03.67.Lx, 03.65.Db, 04.60.-m, 95.35.+d, 04.90.+e

## 1. Introduction

In a previous paper [1], a study of the properties of gravitational eigenstates was begun, the motivation for which originated from earlier work [2] which suggested that certain members of the quantum eigenstate ensemble of deep gravitational wells might have properties that may make them relevant for understanding the nature and origin of some astrophysical phenomena, such as Dark Matter. The motivation for further theoretical study of gravitational eigenstates is made even more compelling by the recent experimental verification of their existence [3, 4].

In particular, paper [1] developed general, non-integral formulae for calculating the transition rates of state-to-state dipole-allowed transitions within the eigenspectra of gravitational potential wells. Sometimes these non-integral formulae continue to have limitations caused by a large, non-truncatable number of summation terms, and/or excessively large numbers, and other more approximate methods need to be employed. The purpose of this paper therefore is to develop further practical methods for these cases which are of use in approximating matrix elements for both the spontaneous dipole decay of the high-quantum-number gravitational eigenstates, and potentially also of more general interactions. In particular we examine ways of determining the approximate overall position, size and internal structure of the individual eigenstates, in particular their radial components.

The position, shape and structure of the eigenstates turn out to be very useful in determining certain dipole transition overlap integrals and can aid in setting maximum limits on the absolute value of others. They are also important in investigating limits on higher order (multi-pole) decays and overlap integrals for other more general interactions, with both electromagnetic radiation and particles. (For example, it is clear that if two states share no common region of space then no interaction Hamiltonian could induce transitions between them.) Structural features of the wavefunction components of an overlap integral can also strongly influence its value even when the states do share a significant spatial overlap region. Indeed there are particular individual eigenstates whose structure and position are such that the only significant transitions possible are those that have a negligible effect on the perturbing entity.

Important structural features of eigenfunctions include variation of the average absolute value of the wavefunction across its spatial range, and the spatial oscillation frequency (SOF) of the zeros or nodes of the eigenfunction components. We wish to establish how these features vary with the location of the state on the quantum state diagram. This diagram is reproduced here from [1] as figure 1, and uses the quantum parameter $p\,(\equiv n-l)$ as defined in that paper.

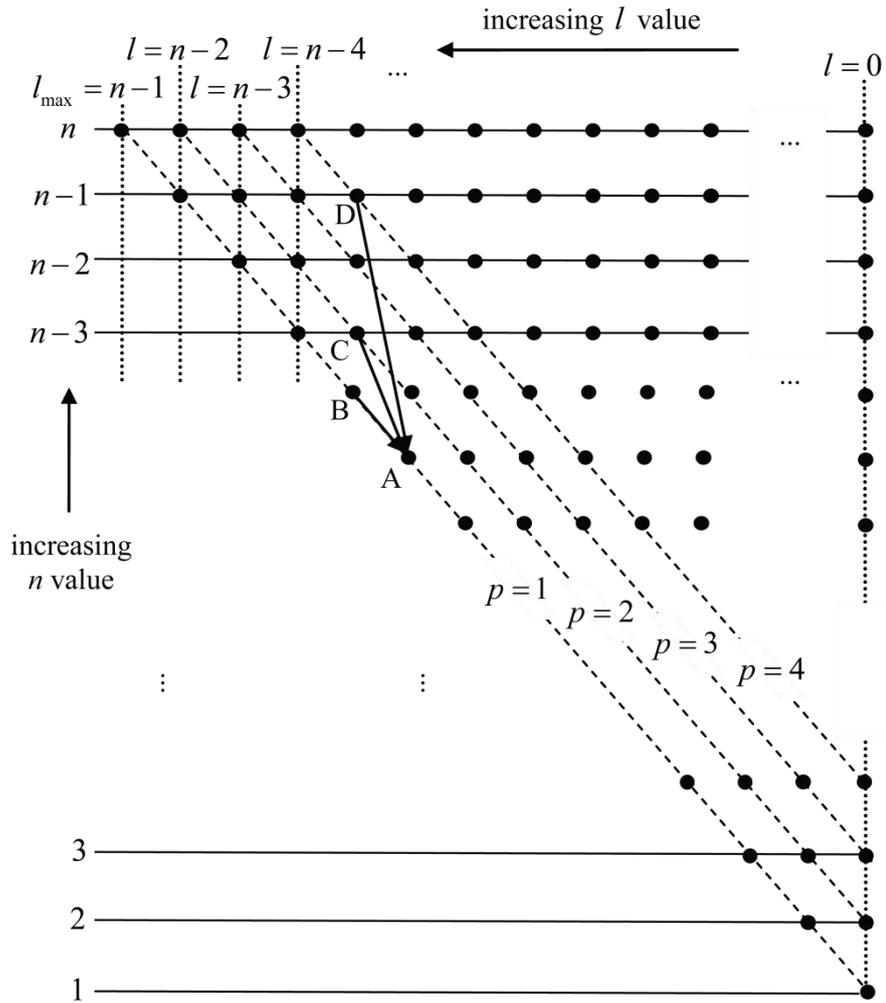

**Figure 1.** Schematic representing the high $n$, $l$, $m$ -valued stationary states, drawn to emphasize the parameter $p \equiv n - l$. Each solid circle on the diagram represents $2l + 1$ $z$ -projection substates.

More specifically, for a given eigenstate on figure 1, we wish to know, how far away on figure 1 a second state must lie (in terms of changes in the quantum parameters $\Delta n$, $\Delta l$ and/or $\Delta p$) so that differences in the structural features of the two states will ensure a negligible probability interaction. This will depend on the location of the initial state and on the direction of the second state.

In section 2 we consider the effect that the size and shape of the angular component wavefunctions have on the value overlap integral, particularly in the limiting case of large quantum numbers, and derive limiting values for the angular component of the dipole matrix element. Section 3 deals with the corresponding radial component of the dipole matrix element. We firstly look at exact results for very low-$p$ that demonstrate the rapid reduction in the radial overlap component for high-$n$ as the difference between the quantum state initial and final $p$ values $\Delta p$ and corresponding principle quantum number change $\Delta n$ increases. We then examine a procedure developed to determine the approximate positions of the zeros of the radial eigenstate wavefunctions, which enables the width, spatial oscillation frequency and average position of any radial eigenfunction to be obtained for arbitrary values of $n$ and $l$. Combining the results from section 2 and 3 then enables decay rates for various significant quantum states to be obtained and this is done in section 4. Section 4 also deals with calculating the decay rates of other constant-$p$ transitions, showing that for low-$p$, the rates are similar to those for the $p = 1$ states. In section 4.3 transitions involving $\Delta n = 2, 3, 4 \ldots$, but those that do not necessarily end on $p = 1$ are examined, and procedures and formulae for dealing with the low-$p$ states with multiple decay channels is developed. This section then uses some of the methods developed to obtain trends in decay rates as one moves vertical and horizontally away from the $p = 1$ diagonal of figure 1, demonstrating how the long lifetimes of states decrease as one moves away from the low-$p$ diagonals. We conclude with a summary and discussion of the salient points of the paper in section 5.

## 2. The effect of the angular eigenfunction components on $\Pi_{if}$

We consider separately the spherical harmonic (angular) and associated Laguerre (radial) eigenfunction components. As in [1] the total dipole decay rate is given by

$$A_{i,f} = \frac{\omega_{if}^{3} \left| \left\langle f \left| e\mathbf{r} \right| i \right\rangle \right|^{2}}{3\varepsilon_{0}\pi\hbar c^{3}} = \frac{\omega_{if}^{3}\Pi_{if}^{2}}{3\varepsilon_{0}\pi\hbar c^{3}} \qquad (1)$$

where as in [1] $e$ is the electronic charge, $\varepsilon_{0}$ the electrical permittivity of free space, $\left| \left\langle f \left| e\mathbf{r} \right| i \right\rangle \right| = \Pi_{if}$ the absolute value of the dipole matrix element for spontaneous decay for the transition $i$ to $f$, $\omega_{if}$ the angular frequency corresponding to the transition $i$ to $f$, $\mu$ the reduced mass, and the other symbols have their usual meanings. $\Pi_{if}$ is given again as in [1] by

$$\Pi_{if} = \left| \int_0^\infty \int_0^\pi \int_0^{2\pi} R_{nf,lf}^* \, Y_{lf,mf}^* \, e\mathbf{r} \, R_{ni,li} \, Y_{li,mi} \, r^2 \sin(\theta) \, d\phi \, d\theta \, dr \right|$$

$$= \sqrt{\left( \Pi_{ifx}{}^2 + \Pi_{ify}{}^2 + \Pi_{ifz}{}^2 \right)} \tag{2}$$

where

$$\Pi_{ifx} = e \int_0^\infty R_{nf,lf}^* \, r^3 \, R_{ni,li} \, dr \int_0^\pi \int_0^{2\pi} Y_{lf,mf}^* \, Y_{li,mi} \sin(\theta)\cos(\phi)\sin(\theta) d\phi d\theta$$

$$\equiv e \, I_R \, I_{\theta\phi x}$$

$$\Pi_{ify} = e \int_0^\infty R_{nf,lf}^* \, r^3 \, R_{ni,li} \, dr \int_0^\pi \int_0^{2\pi} Y_{lf,mf}^* \, Y_{li,mi} \sin(\theta)\sin(\phi)\sin(\theta) d\phi d\theta$$

$$\equiv e \, I_R \, I_{\theta\phi y}$$

$$\Pi_{ifz} = e \int_0^\infty R_{nf,lf}^* \, r^3 \, R_{ni,li} \, dr \int_0^\pi \int_0^{2\pi} Y_{lf,mf}^* \, Y_{li,mi} \cos(\theta)\sin(\theta) d\phi d\theta$$

$$\equiv e \, I_R \, I_{\theta\phi z}$$

$$\text{and } I_R = \int_0^\infty R_{nf,lf}^* \, r^3 \, R_{ni,li} \, dr \, ,$$

$$I_{\theta\phi x} = \int_0^\pi \int_0^{2\pi} Y_{lf,mf}^* \, Y_{li,mi} \sin(\theta)\cos(\phi)\sin(\theta) d\phi d\theta \, ,$$

$$I_{\theta\phi y} = \int_0^\pi \int_0^{2\pi} Y_{lf,mf}^* \, Y_{li,mi} \sin(\theta)\sin(\phi)\sin(\theta) d\phi d\theta \, \text{ and}$$

$$I_{\theta\phi z} = \int_0^\pi \int_0^{2\pi} Y_{lf,mf}^* \, Y_{li,mi} \cos(\theta)\sin(\theta) d\phi d\theta$$

with the $R_{ni,li}$ being related to the associated Laguerre polynomials and $Y_{li,mi}$ the spherical harmonics.

### 2.1. The effect of shape and spread of the probability density distribution on $\Pi_{if}$

The azimuthal ($\phi$) dependence of all spherical harmonic eigenfunctions components is of the form $e^{\pm im\phi}$ so that the probability density in the azimuthal direction is uniform for all states, and hence there is always functional overlap in this coordinate direction. The spherical harmonic polar ($\theta$) dependence varies with the values of both $l$ and $m$. A zero angular momentum $z$-projection value (i.e. $m = 0$) corresponds to the highest degree of polar angular spread for a given $l$ value, with the lobes of probability density equally spaced around the polar direction (figure 2, curve ($f$)). A state with a high $z$-component of angular momentum ($m$ close to $l$) however corresponds to a polar distribution that is more flattened and disk-like, but with the probability density lobes still approximately equally spaced and centred around $\theta = \pi$ (figure 2($a$)). Clearly for any two spherical harmonics, there will always be some degree of overlap in the polar direction as well and the effect that this has on $I_{\phi\theta x}$, $I_{\phi\theta y}$ and $I_{\phi,\theta,z}$ in the limit of large $l$ needs to be determined. It also means that two eigenstates can be spatially disjoint only if the radial components of their wavefunctions are physically separated in space. We will return to the radial components after examining the limiting values of the angular part of the dipole transition overlap integral.

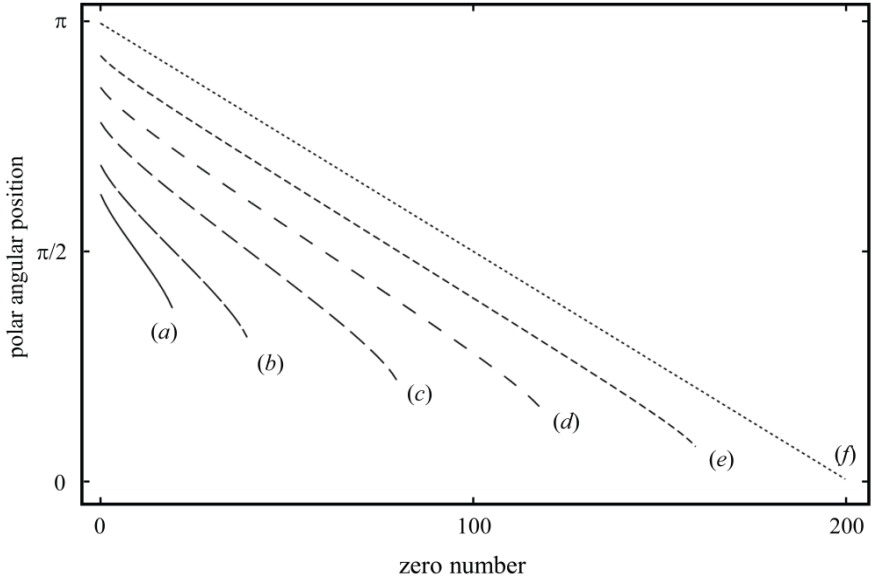

**Figure 2.** Polar angular position (0 to $\pi$) of the zeros of the probability density functions of the spherical harmonics as a function of zero number for $l = 200$ and various $m$ values. (*a*) $m = 180$; (*b*) $m = 160$; (*c*) $m = 120$; (*d*) $m = 80$; (*e*) $m = 40$; (*f*) $m = 0$;

### 2.2. Limiting values of the angular components of the overlap integral

Unlike $I_R$, which can become arbitrarily large because of the presence of the extra factor of $r$, the integrals $I_{\theta\phi x}$, $I_{\theta\phi y}$ and $I_{\theta\phi z}$ are limited in size even when there is considerable overlap. Since we are examining states where the values of $n$ are large but $p$ is small, we will generally be concerned with determining how the size of these integrals behaves when the initial value of $l$ is large. The initial values of $m$ are arbitrary however and each of the points on the low-$p$ diagonals of figure 1 represents all allowed values of $m$ from $-l$ to $+l$. There are essentially two limiting regimes. Either $m$ remains small as $l \to \infty$ or both $l$ and $m \to \infty$ with $l \sim m$. Table 1 shows the limiting behaviour of angular components $I_{\theta\phi x}$, $I_{\theta\phi y}$ and $I_{\theta\phi z}$, (a) when $m$ is small compared to $l$ and $(l-m) \to \infty$, and (b) when $|m|$ is near $l$ and $l, m \to \infty$. Dipole transitions are accompanied by a angular momentum change $\Delta l = \pm 1$ and a $z$-projection change of $\Delta m = \pm 1$.

**Table 1.** Limiting values for the angular component of the dipole matrix elements for electromagnetic decay: (a) $\lim(l-m\to\infty)$, when $m \ll l$ and (b) $\lim(l, m \to \infty)$ when $m \sim l$.

(a)

| | | | $\Delta m$ | | | |
|---|---|---|---|---|---|---|
| | | | +1 | | 0 | −1 | |
| | | | $I_{\theta\phi x}$ | $I_{\theta\phi y}$ | $I_{\theta\phi z}$ | $I_{\theta\phi x}$ | $I_{\theta\phi y}$ |
| $\Delta l$ | +1 | $m > 0$ | $-\tfrac{1}{4}$ | $\tfrac{i}{4}$ | $\tfrac{1}{2}$ | $\tfrac{1}{4}$ | $\tfrac{i}{4}$ |
| | −1 | | $\tfrac{1}{4}$ | $-\tfrac{i}{4}$ | $\tfrac{1}{2}$ | $-\tfrac{1}{4}$ | $-\tfrac{i}{4}$ |
| | +1 | $m = 0$ | $-\tfrac{1}{4}$ | $\tfrac{i}{4}$ | $\tfrac{1}{2}$ | $\tfrac{1}{4}$ | $\tfrac{i}{4}$ |
| | −1 | | $\tfrac{1}{4}$ | $-\tfrac{i}{4}$ | $\tfrac{1}{2}$ | $-\tfrac{1}{4}$ | $-\tfrac{i}{4}$ |
| | +1 | $m < 0$ | $-\tfrac{1}{4}$ | $\tfrac{i}{4}$ | $\tfrac{1}{2}$ | $\tfrac{1}{4}$ | $\tfrac{i}{4}$ |
| | −1 | | $\tfrac{1}{4}$ | $-\tfrac{i}{4}$ | $\tfrac{1}{2}$ | $-\tfrac{1}{4}$ | $-\tfrac{i}{4}$ |

(b)

| | | | $\Delta m$ | | | |
|---|---|---|---|---|---|---|
| | | | +1 | | 0 | −1 | |
| | | | $I_{\theta\phi x}$ | $I_{\theta\phi y}$ | $I_{\theta\phi z}$ | $I_{\theta\phi x}$ | $I_{\theta\phi y}$ |
| $\Delta l$ | +1 | $m > 0$ | $-\tfrac{1}{2}$ | $\tfrac{i}{2}$ | $0$ | $0$ | $0$ |
| | −1 | | $0$ | $0$ | $0$ | $-\tfrac{1}{2}$ | $-\tfrac{i}{2}$ |
| | +1 | $m < 0$ | $0$ | $0$ | $0$ | $\tfrac{1}{2}$ | $\tfrac{i}{2}$ |
| | −1 | | $\tfrac{1}{2}$ | $-\tfrac{i}{2}$ | $0$ | $0$ | $0$ |

### 2.3. The effect of limiting values of the angular components of the overlap integral on eigenstructure shape

The values in table 1 can be used to predict trends that might occur when upward and downward transitions take place (for example due to stimulated emission and absorption that could occur in instances where photon densities were sufficiently high over the whole of the eigenstate's volume). For states whose *z*-projection angular momentum value *m* is close to zero (table 1(a)), both upward and downward changes in *m* are equally likely. If sufficient radiation density were present to result in significant numbers of stimulated transitions then the effect of this would be to diffuse the state population evenly around *m* = 0. However for those states that have an absolute value of *m* close to *l*, transitions take place in a manner better described by table 1(b). This shows that for transitions where *l* is increased, states with positive *m* have their *m* value further increased and states with negative *m* values have their *m* values further decreased to more negative values. Thus transitions to larger *l* favour a net migration towards states with an increased *z*-component of angular momentum, (more 'oblate'). Conversely transitions to lower *l* are accompanied by a net migration toward states of lower z-component of angular momentum, that is, more highly spherical states.

In the atomic case there is no significant correlation of change of *n* value with change of *l*. However as will be seen later, in high-*n* states crossing *p* diagonals in a transition has a large effect on the value of $I_R$, and increases and decreases in *l* are correlated with respective increases and decreases in *n*. (That is for example, a $\Delta l$ of +1 is much more likely to be accompanied by a $\Delta n$ of +1 rather than a $\Delta n$ of −1, for high *n* state.) Then, since state size is related to the *n* value, an eigenstructure containing many particles might be expected to exhibit differences in shape that reflect the degree of stimulated emission and absorption that has taken place in its history. A randomly populated eigenstructure ought to be spherical, since the sum over all *m* states of any fixed total angular momentum value *l* is spherically symmetric. If however an eigenstructure was subject to significant stimulated transfers in its past it might have an inner structure that is more spherical but an outer structure that could be somewhat oblate or flattened.

### 3. Radial component

We now return to the radial components of the eigenfunctions and the radial dipole overlap integral component $I_R = \int_0^\infty R^*_{nf,lf} \, r^3 \, R_{ni,li} \, dr$ which is of most importance in determining the radiative lifetime of any high quantum number state. We recall the significance of the $p$ value on the state diagram of figure 1, that is that it reflects the spatial oscillatory behaviour of the eigenfunction ($p = 1$ corresponding to a radial eigenfunction with one peak, no zeros; $p = 2$, two-peaked, one-zero; etc.) We are interested in the "beginning" and "ending" position of the eigenfunctions as well as their total radial spatial extent and their spatial oscillation frequencies because these things have the most profound effect on determining the value of the overlap integrals. The radial wavefunction components have the greatest radial extent when $l = 0$, from low $r$ up to $r \sim 2n^2 b_0$ where $b_0$ is a scale factor [2].

### 3.1. Exact calculation of $I_R$ and eigenfunction position, size and spread, for low-p eigenstates, and the demonstration of rapid decay

When the $l$ value is very close to $n$, the Laguerre polynomials contain only a small number of terms (determined by the value of $p = n - l$), and the physical structure and overlap integrals can be examined exactly. We define the width of a $p = 1$ eigenfunction as the full width at zero concavity (FWZC). The peak centre and FWZC of a $p = 1$ ($R_{n,l=n-1}$) eigenfunction are given by $n(n-1)b_0 \sim n^2 b_0$ and $2n\sqrt{n-1}\,b_0 \sim 2n^{3/2}b_0$ respectively. This means that two $p = 1$ radial eigenfunctions need differ in $n$ values by $\sim \sqrt{n}$ before their radial position is significantly different compared to their width. For larger separations the amplitude of either function at the position of the peak of the other drops very rapidly. It may be shown that given two $p = 1$ eigenfunctions whose peaks are separated by $w$ FWZCs, that the ratio of the amplitude of one wavefunction at the position of the peak of the other, to the amplitude of the peak of the first is given by

$$\exp\left[-2w\sqrt{n_i-1}\right]\left(\frac{n_i + 2w\sqrt{n_i-1}}{n_i}\right)^{n_i-1} \sim \exp\left[-2w\sqrt{n_i}\right]\left(1+\frac{2w}{\sqrt{n_i}}\right)^{n_i-1} \qquad (3)$$

and that the radial dipole overlap integral $I_R$ has an approximate value of

$$b_0 \sqrt{\frac{2(n_i n_f)^{5/2}}{n_i + n_f}} \left(\frac{n_f}{n_i}\right)^{n_i - n_f} \qquad (4)$$

where $n_i$ and $n_f$ are the ($\gg 1$) initial and final quantum levels.

The dramatic behaviour of (3) and (4) for large $n$ is demonstrated in the following example. For a central mass of $10^{30}$ kg with $n = 10^{23}$ (binding energy $\sim 2$ eV at a radius of $\sim 6 \times 10^{11}$ m for a proton mass), the $p = 1$ state has only one possible downward dipole-allowed transition: $\Delta p = 0$, $\Delta n = 1$. The initial and final states exhibit virtually 100% overlap but an angular emission frequency of less than 0.1 $\mu$Hz, so that the dipole decay rate is negligible ( $A_{i,f} = \dfrac{\omega_{if}{}^3 \Pi_{if}{}^2}{3\varepsilon_0 \pi h c^3}$ ).

Multi-pole electromagnetic transitions and other interactions can however allow $\Delta n > 1$ transitions so the degree of separation of two $\Delta p = 0$ states for arbitrary $\Delta n$ becomes an important consideration for determining the maximum value of an overlap integral in these cases. Taking the dipole decay as an example however, as $\Delta n$ increases the overlap gradually decreases and so does $I_R$. For $w = 1$, (3) gives the amplitude ratio as 0.4, but for $w = 10$ (corresponding to a transition with $\Delta n \sim 3 \times 10^{12}$) it is $1.4 \times 10^{-87}$ and for $w = 100$ it is $\sim 1 \times 10^{-8686}$. The corresponding $I_R$ values are $\sim 3 \times 10^{11}$ for $w = 1$, $\sim 1 \times 10^{-30}$ for $w = 10$ and $\sim 1 \times 10^{-4162}$ for $w = 100$. The overwhelming decrease in the corresponding overlap integral $I_R$ when the two wavefunctions are spatially distinct far outweighs increases in $A_{i,f} = \dfrac{\omega_{if}{}^3 \Pi_{if}{}^2}{3\varepsilon_0 \pi h c^3}$ that might result from increases in other parameters due to larger $\Delta n$, such as $\omega_{if}$ in this case (for $w = 10$, $\omega_{if}$ increases to $\sim 50$ kHz). This clearly demonstrates the effective extinction of interaction between two $p = 1$ states as $w$ increases.

For larger $n$ the effect is even more dramatic: for a $10^{42}$ kg central mass potential, a transition that takes place from a $n = 10^{34}$, $p = 1$ eigenstate, requires a quantum level change of $\Delta n \sim 3 \times 10^{21}$ to produce a 50kHz photon. However the degree of overlap for this value of $\Delta n$ is such that the radial component of the dipole matrix element $I_R = \int_0^\infty R^*_{nf,lf} \, r^3 R_{ni,li} \, dr$ has a value of only $\sim 10^{-3.7 \times 10^8}$. A similar analytical procedure can be undertaken for $p = 2, 3, 4...$ with similar results except that, because the states have a progressively larger spatial extent, the required value of $\Delta n$ to ensure distinct separation of states in space is now be larger. This concept is illustrated diagrammatically in figure 3.

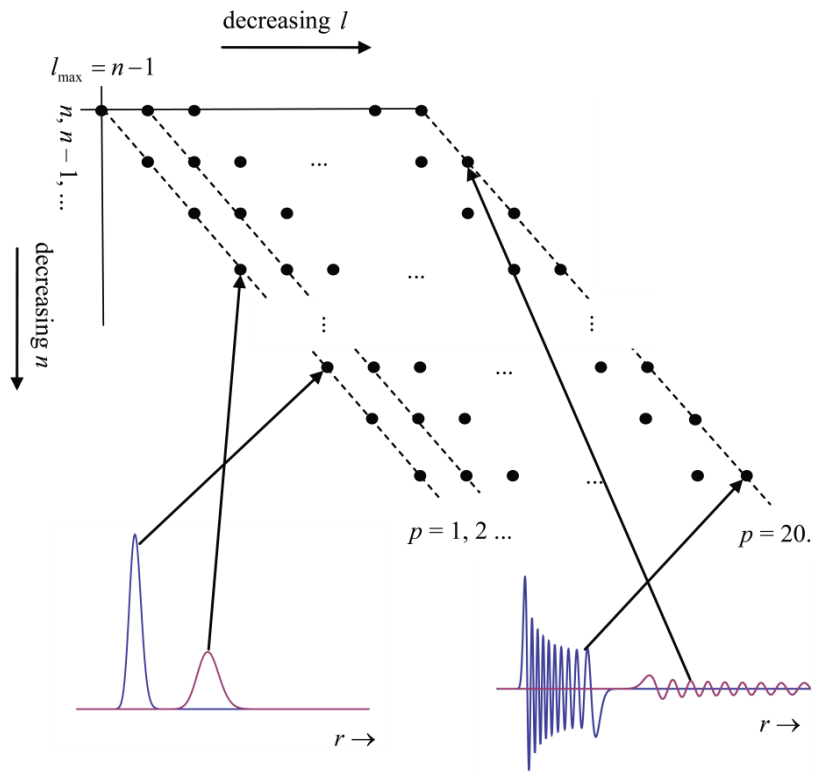

**Figure 3.** Schematic diagram showing changes in position of high-*n*, low-*p*, radial component eigenfunctions with *n*, and the significance of *p* in determining the change in *n* required for radial distinct radial separation of the eigenfunctions.

### 3.2. Position, width and zeros of the high-n radial eigenfunctions

The importance of the width, position and SOF of any given eigenfunction means that it is useful to have a reliable way to determine these structural parameters. We now develop an equation for determining the position, extent and zeros of the radial eigenfunctions. Figure 4 shows the typical oscillatory form of the radial function $R_{n,l}$ $r$ and diagrammatically shows the way that the terms "first zero spacing", last zero spacing" and "estimated polynomial width" are defined. The SOF is defined as the reciprocal of the zero spacing at any position.

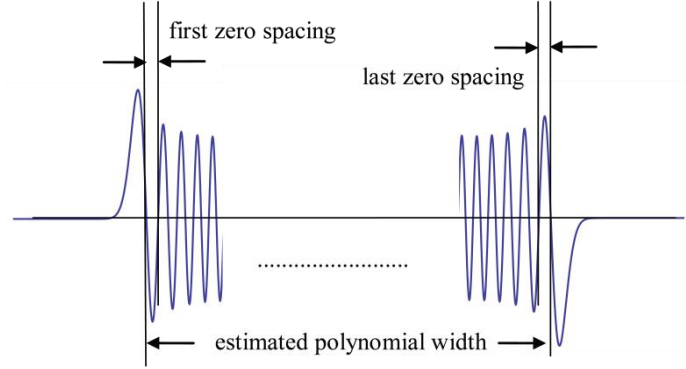

**Figure 4.** Schematic diagram showing the typical oscillatory behaviour of the function $R_{nl}(r)\,r$, the first and last zero spacing and the criteria for estimating the Laguerre polynomial width.

$R_{n,l}\,r$, written in terms of $p$, is

$$\left(\left(\frac{2}{n_i b_0}\right)^3 \left(\frac{(p-1)!(2n_i-p)!}{2n_i}\right)\right)^{\frac{1}{2}} \exp\left(-\frac{r}{n_i b_0}\right)\left(\frac{2r}{n_i b_0}\right)^{(n_i-p)} r \times$$
$$\left(\sum_{k_i=0}^{(p-1)} \frac{(-1)^{k_i}\left(\frac{2r}{n_i b_0}\right)^{k_i}}{(p-k_i-1)!(2n_i-2p+k_i+1)!k_i!}\right). \quad (5)$$

A little algebra shows that this function, and also the functions $R_{n,l}$ and $\left(R_{n,l}\,r\right)^2$, have the same zeros as the polynomial:

$$\Upsilon(\gamma) = \sum_{k=0}^{p-1} \frac{(-1)^{k-p+1}(\gamma)^k (p-1)!(2n-p)!}{(p-1-k)!(2n-2p+k+1)!k!} \quad (6)$$

where $\gamma = 2r/(nb_0)$ and again $p = n - l$. Once the properties of $\Upsilon(\gamma)$ are determined, it is then a straightforward matter to obtain the properties of $R_{n,l}$, $R_{n,l}\,r$, $\left(R_{n,l}\,r\right)^2$ etc.

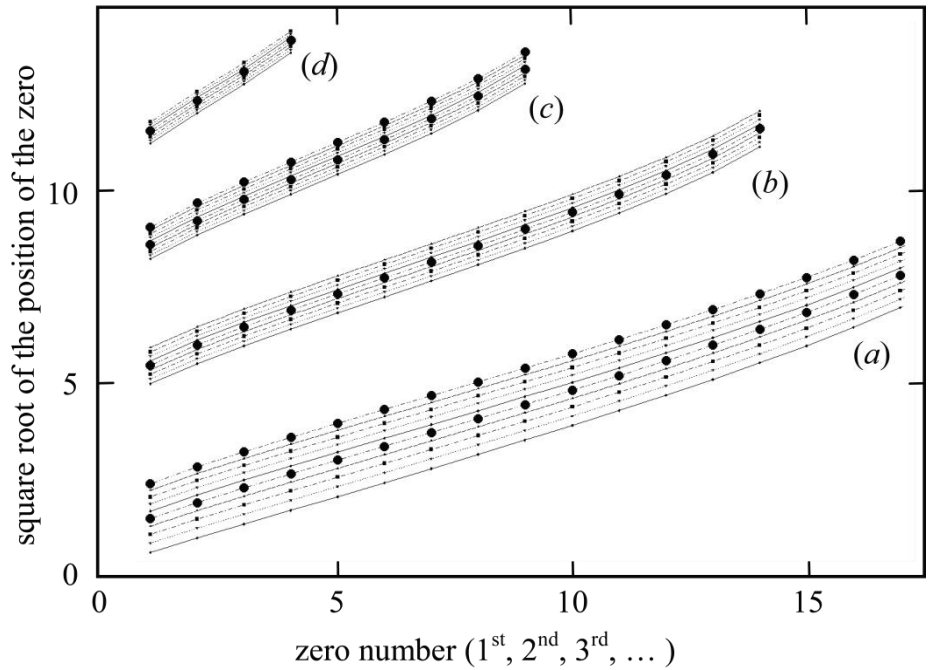

**Figure 5.** Square roots of the 1$^{st}$, 2$^{nd}$, 3$^{rd}$, ... $i^{th}$, ... zeros of the function

$\Upsilon(\gamma) = \sum_{k=0}^{p-1} \frac{(-1)^{k-p+1}(\gamma)^k (p-1)!(2n-p)!}{(p-1-k)!(2n-2p+k+1)!k!}$ as a function of the zero number.

Curve set ($a$): $p = 20$, $n = \{21,...30\}$; curve set ($b$): $p = 15$, $n = \{41,...50\}$; curve set ($c$): $p = 10$, $n = \{61,...70\}$; curve set ($d$): $p = 5$, $n = \{81,...89\}$.

We wish to obtain from (6) estimates of the spatial extent and position of $\Upsilon(\gamma)$, the spacing (spatial periods) of its zeros, and the height of its peaks. The process begins with an empirical observation shown in figure 5 that the set of square roots of the zeros of $\Upsilon(\gamma)$, plotted as a function of zero number, has a consistent characteristic form that is approximately linear. The degree to which linearity is approached is independent of the value of $p$ or $n$ separately but is better when $p \ll n$ and $n \gg 1$ concurrently (curve set ($d$), figure 5). $\Upsilon(\gamma)$ is a polynomial of degree $p-1$ and may be written in terms of its $p-1$ zeros, $\gamma_1, \gamma_2, \gamma_3, \dots \gamma_{p-1}$       as       $\Upsilon(\gamma) = (\gamma - \gamma_1)(\gamma - \gamma_2)(\gamma - \gamma_3)\dots(\gamma - \gamma_{p-1})$.
Mathematically, if the square roots of the zeros of $\Upsilon(\gamma)$ are linearly related, then the $i^{\text{th}}$ zero $\gamma_i$ may be written in terms of the first zero $\gamma_1$ as

$$\sqrt{\gamma_i} = \sqrt{\gamma_1} + (i-1)\delta \tag{7}$$

where $\delta$ is a 'spacing' parameter that is constant for a given value of *n* and *p*. We let the sum of the zeros be $\alpha = \sum_{i=1}^{p-1} \gamma_i$ and the sum of the products of the zeros taken two at a time be $\beta = \sum_{i,j=1,i\neq j,i>j}^{p-1} \gamma_i \, \gamma_j = \frac{1}{2}\left( \sum_{j=1}^{p-1}\sum_{i=1}^{p-1} \gamma_i \, \gamma_j - \sum_{i=1}^{p-1} \gamma_i^2 \right)$.
Substituting $\sqrt{\gamma_i} = \sqrt{\gamma_1} + (i-1)\delta$ and $\sqrt{\gamma_i} = \sqrt{\gamma_1} + (i-1)\delta$ into these summation expressions for $\alpha$ and $\beta$ and simplifying gives:

$$\alpha = \frac{1}{6}(p-1)\left( \left(2p^2 - 7p + 6\right)\delta^2 + 6(p-2)\delta\sqrt{\gamma_1} + 6\gamma_1 \right) \tag{8}$$

$$\beta = \frac{1}{360}(p-1)(p-2)\begin{pmatrix} \left(20p^4 - 156p^3 + 427p^2 - 477p + 180\right)\delta^4 + \\ 120\left(p^3 - 6p^2 + 11p - 6\right)\delta^3\sqrt{\gamma_1} + \\ 60\left(5p^2 - 20p + 18\right)\delta^2\gamma_1 + \\ 360(p-2)\delta\gamma_1^{3/2} + 180\gamma_1^2 \end{pmatrix} \tag{9}$$

From simple polynomial theory $-\alpha$ is the coefficient of $\gamma^{p-2}$ and $\beta$ the coefficient of $\gamma^{p-3}$ in equation (6). Using equation (6) gives

$$\alpha = (2n-p)(p-1) \tag{10}$$

$$\beta = (2n-p)(2n-p-1)(p-1)(p-2)/2 \tag{11}$$

Eliminating $\alpha$ and $\beta$ using equations (8 - 11) gives:

$$\frac{1}{6}\left( \left(2p^2 - 7p + 6\right)\delta^2 + 6(p-2)\delta\sqrt{\gamma_1} + 6\gamma_1 \right) - (2n-p) = 0 \tag{12}$$

$$180(2n-p)(2n-p-1) = \begin{pmatrix} \left(20p^4 - 156p^3 + 427p^2 - 477p + 180\right)\delta^4 + \\ 120\left(p^3 - 6p^2 + 11p - 6\right)\delta^3\sqrt{\gamma_1} + \\ 60\left(5p^2 - 20p + 18\right)\delta^2\gamma_1 + \\ 360\left(p-2\right)\delta\,\gamma_1^{3/2} + 180\gamma_1^2 \end{pmatrix} \quad (13)$$

Eliminating $\delta$ from (12) and (13) gives two polynomial equations in $\sqrt{\gamma_1}$. Some lengthy algebra shows these polynomial equations yield the same quartic equation in $\gamma_1$ which simplified gives:

$$\begin{pmatrix} 4(p-2n)^2(2p-3)\left(30 + p\left(-50 + (37-9p)p + n(8p-19)\right)\right)^2 \\ +24p(p-2n)^2\left(210 + p\begin{pmatrix} -380 + n\left(191 + p\left(-303 + (155-26p)p\right)\right) \\ + p\left(117 + p\left(139 + 2p(9p-50)\right)\right) \end{pmatrix}\right)\gamma_1 \\ +4(2n-p)p\left(-90 + p\begin{pmatrix} -(27+4p(2p-7))(-10+p(9p-16)) + \\ n\left(-669 + 2p\left(596 + p(62p-335)\right)\right) \end{pmatrix}\right)\gamma_1^2 \\ +12p^2(p-2n)(2p-1)\left(11 + p(3p-11)\right)\gamma_1^3 + (1-2p)^2p^2(2p-3)\gamma_1^4 \end{pmatrix} = 0$$

$$(14)$$

Although not repeated here because they are rather lengthy, it is immediately possible to solve (14) and hence write down formulae for the approximations to the first ($\gamma_1$) and last ($\gamma_{p-1}$) zeros of $\Upsilon(\gamma)$, and the spacing parameter $\delta$, directly in terms of any value of $n$ and $p$. Using (7) then enables approximations of all of the zeros of $\Upsilon(\gamma)$ to be obtained, thus enabling estimation of the spread and extent of $R_{n,l}$ and the separation $\Delta r_i = (\gamma_i - \gamma_{i-1})n\,b_0/2$ of each of its zeros.

Approximations are excellent for values of $n$ and $p$ that can be checked by direct numerical solution of the relevant polynomial. For example, even when the linearity condition, $p \ll n$ and $n \gg 1$ is only marginally met, as when $n = 30$ and $p = 10$, the approximation formulae zero set is {22.0, 27.6, 33.7, 40.5, 47.9, 56.0, 64.6, 73.9, 83.8} while the direct numerical solution zero set of $\Upsilon(\gamma)$ is {22.4, 28.3, 34.3, 40.5, 47.3, 54.8, 63.2, 73.2, 85.9}, that is accuracies of $\sim 5\%$ or less. However when the linearity condition $p << n$ and $n >> 1$ is well satisfied, the degree of accuracy of zero position approximation is remarkable. This can be seen in the following. For small sufficiently small values of $p$, the eigenfunctions may be exactly calculated analytically even when $n >> 1$, since the radial eigenfunction component contains only $p$ terms. These analytically obtained values may then be compared with those obtained using the empirical relationship (7). With $n = 1.0 \times 10^{34}$ and $p = 4$, the present square root approximation treatment here gives the positions of the zeros of $R_{n,l}$ (which are located at a radius $r$ of $\sim 3.6 \times 10^{21}$ m) to better than 1 part in $10^{18}$ when compared with the numerical calculation of the zeros using the exact formula for $R_{n,n-4}$. This leads to the accuracy of the actual spacing between the zeros (in this case $\sim 44$ km) being better than 4% even at these vast distances.

### 3.3. Significance of the zeros of the radial eigenfunctions for overlap integral values

When the initial and final wavefunctions do overlap, there are other reasons why the overlap integrals may still remain negligibly small in the case of large $n$. In dipole transitional decays originating from a 'deep' state ($p >> 1$) and ending on a low $p$ state, such as D to A of figure 1 where there is only unit change in $\Delta l$, the two states usually still have extensive overlap despite a large change in $n$ because the much larger value of $p$ in the initial state means that it can be spread over a large radial extent. One clear reason for a low value of $I_R$ in this case is that the normalisation condition on the upper eigenfunction requires that its average amplitude is very small because the function is highly spread. Another reason concerns the relative spatial oscillation frequencies (RSOF's) of the two states, particularly that of the high-$p$, upper state, which we now discuss.

The preceding section enables us to determine the spacing of the zeros in the upper level and compare them with the width of the $p = 1$ level. Considering again a central mass of $10^{42}$ kg, the use of equation (14) shows that for $\Delta n = 10^{23}$ (a proton quantum number change necessary to result in emission of a 0.3 MHz photon) the spacing of those zeros in the higher-$n$ eigenfunction that lie under the profile of the initial eigenstate is $\sim 2 \times 10^{-7}$ m, and constant (due to the particular value of the spacing parameter in these cases) over the whole of the radial eigenfunction at A in figure 1 to more than 1 part in $10^5$. This means that there are over $\sim 4 \times 10^{11}$ uniform oscillations of the initial wavefunction under the relatively symmetric $p_f = 1$ final state function profile. The cancellation effect of this oscillatory behaviour on the radial dipole integral can be seen by noting the approximate value of the integral given by the formula

$$I_R = b_0 n_i \left(\frac{2}{\pi p_i}\right)^{\frac{1}{4}} \left(\frac{e}{2}\right)^{\frac{p_i}{2}} \left(\frac{p_i}{n_i}\right)^{\frac{p_i-3}{2}} = I_R \sim 10^{-1\times 10^{24}} b_0 \text{ when } p_f = 1, \ n_i \approx 10^{34},$$

$\Delta n = p_i \approx 10^{23}$ [1].

## 4. Decay rates

We now examine numerical values of, or upper limits to, the dipole decay rates of some of the gravitational eigenstates located at various positions on figure 1. These numerical values are obtained either by direct integration of the relevant overlap integrals (small $p$), application of the various approximations developed in this series of papers or, in the most difficult cases, orders of magnitude estimated from the degree (or lack thereof) of overlap of the component eigenfunctions and their relative spatial oscillation frequencies.

### 4.1. Decay times for states along the $p = 1$ diagonal ($\Delta n = 1$)

The transitions involving only states along the $p = 1$ diagonal with high-$n$ values have directly integrable radial overlap integrals, and potentially have the longest radiative lifetimes since only one dipole decay channel is available, the adjacent $\Delta n = 1$, $p = 1$ state (B to A in figure 1). Table 2 gives dipole decay times for some of these $\Delta n = 1$ transitions for various central masses and relevant, suitably sized quantum numbers. The values are 'representative only' of the various transitions, in the sense that a nominal value of 0.5 has been taken for the angular component of the dipole matrix element. (Individual transition integrals between the various substates will vary considerably between values of this order and zero.) Their purpose is to give an indication of expected transition rates for various conditions and average upper limits on their values.

**Table 2.** Binding energies, positions, frequencies and decay times for high-$n$, $p = 1$ eigenstates

| Central mass (kg) | Eigenstate particle mass (kg) | Quantum number, $n$ | Energy (eV) | Radius of eigenstate (m) | Emission frequency (Hz) | Dipole decay time (s) |
|---|---|---|---|---|---|---|
| $6.0 \times 10^{24}$ | $1.7 \times 10^{-27}$ | $10^{18}$ | $0.21$ | $1.0 \times 10^{7}$ | $\sim 1 \times 10^{-4}$ | $1.5 \times 10^{15}$ |
| $6.0 \times 10^{24}$ | $9.1 \times 10^{-31}$ | $5.5 \times 10^{14}$ | $1.1 \times 10^{-4}$ | $1.0 \times 10^{7}$ | $\sim 1 \times 10^{-4}$ | $1.5 \times 10^{15}$ |
| $2.0 \times 10^{30}$ | $1.7 \times 10^{-27}$ | $5.8 \times 10^{23}$ | $6.9 \times 10^{-2}$ | $1.0 \times 10^{13}$ | $\sim 6 \times 10^{-11}$ | $\sim 8 \times 10^{21}$ |
| $2.0 \times 10^{30}$ | $9.1 \times 10^{-31}$ | $3.2 \times 10^{20}$ | $3.7 \times 10^{-5}$ | $1.0 \times 10^{13}$ | $\sim 6 \times 10^{-11}$ | $\sim 8 \times 10^{21}$ |
| $2.0 \times 10^{30}$ | $1.7 \times 10^{-27}$ | $1.0 \times 10^{22}$ | $2.3 \times 10^{2}$ | $3.0 \times 10^{9}$ | $\sim 1 \times 10^{-5}$ | $\sim 1 \times 10^{13}$ |
| $2.0 \times 10^{30}$ | $9.1 \times 10^{-31}$ | $5.5 \times 10^{18}$ | $1.2 \times 10^{-1}$ | $3.0 \times 10^{9}$ | $\sim 1 \times 10^{-5}$ | $\sim 1 \times 10^{13}$ |
| $2.0 \times 10^{42}$ | $1.7 \times 10^{-27}$ | $1.0 \times 10^{34}$ | $2.3 \times 10^{2}$ | $3.0 \times 10^{21}$ | $\sim 1 \times 10^{-17}$ | $\sim 1 \times 10^{25}$ |
| $2.0 \times 10^{42}$ | $9.1 \times 10^{-31}$ | $5.4 \times 10^{30}$ | $1.3 \times 10^{-1}$ | $3.0 \times 10^{21}$ | $\sim 1 \times 10^{-17}$ | $\sim 1 \times 10^{25}$ |

*4.2. Other $\Delta n = 1$ transition rates (transitions involving constant $p$)*

States not on the leftmost diagonal involve multiple allowed ($\Delta l = \pm 1$) decay channels and we first consider the primary $\Delta n = 1$ transition decay channel of these states (that is transitions along constant $p$ diagonals). Transition rates for $\Delta n = 1$ transitions along the second leftmost diagonal ($p = 2$) of figure 1 are relatively easily obtained by integration of the wavefunctions and yield similar values to those along $p = 1$. To examine transitions along higher $p$ diagonals we need to specialise the general formula for the radial integral part $I_R$ of $\Pi_{if}$ [1]. $I_R$ (for $l = -1$) may be re-expressed in terms of the final primary quantum numbers $n_f$ and $p_f$ and the size of the quantum level jump $\Delta n$ as

$$I_R = \frac{2^{2n_f - 2p_f + 3}\left(n_f{}^2 + n_f \Delta n\right)^{n_f - p_f + 2} n_f b_0}{\left(2n_f + \Delta n\right)^{2n_f - 2p_f + 5}\left(2n_f - 2p_f + 3\right)!}\left(\frac{\left(p_f - 1\right)!}{\left(p_f + \Delta n - 2\right)!}\right)^{\frac{1}{2}} \times$$

$$\left(\left(2n_f - p_f\right)!\left(2n_f - p_f + \Delta n + 1\right)!\right)^{\frac{1}{2}} \times \qquad (15)$$

$$\sum_{k_f = 0}^{p_f - 1}\left(\begin{array}{l}\dfrac{(2n_f - 2p_f + k_f + 2)(2n_f - 2p_f + k_f + 3)(2n_f - 2p_f + k_f + 4)}{k_f!(p_f - k_f - 1)!} \times \\[2mm] \dfrac{(-2n_f - 2\Delta n)^{k_f}}{(2n_f + \Delta n)^{k_f}}\, {}_2F_1\left(a, b; c; z\right)\end{array}\right)$$

where $\quad {}_2F_1\left(a, b; c; z\right) \quad$ is the hypergeometric function with $a = 2n_f - 2p_f + k_f + 5$, $b = 2 - p_f - \Delta n$, $c = 2n_f - 2p_f + 4$ and $z = \dfrac{2n_f}{2n_f + \Delta n}$.

For the specialised case where $\Delta n = 1$ and $p_f > 2$, this may be written as

$$I_R = \frac{(1 + 2n_f)b_0}{8\left(n_f + 1\right)}\left(\frac{n_f\left(n_f + 1\right)}{\left(n_f + \frac{1}{2}\right)^2}\right)^{n_f - p_f + 3}\sqrt{\left(2n_f - p_f + 1\right)\left(2n_f - p_f + 2\right)} \times$$

$$\prod_{i=2}^{p_f - 2}\left(2n_f - 2p_f + i + 2\right) \times \qquad (16)$$

$$\sum_{k_f = 0}^{p_f - 1}\left(\begin{array}{l}\dfrac{(2n_f - 2p_f + k_f + 2)(2n_f - 2p_f + k_f + 3)(2n_f - 2p_f + k_f + 4)}{k_f!(p_f - k_f - 1)!} \times \\[2mm] \dfrac{(-2n_f - 2)^{k_f}}{(2n_f + 1)^{k_f}}\, {}_2F_1\left(a, b; c; z\right)\end{array}\right)$$

Provided $p\left(= p_f\right) \ll n_f$ and $n_f \gg 1$ we expect that the decay times for all $\Delta n = 1$ transitions along the lines of constant $p = 2, 3, 4...$ will be similar since the overlap is essentially 100% and their radial extent is limited, even though each successive pair of states has a different shape to the $p = 1$ pair (i.e. we expect that the A to B and E to F transitions of figure 1 will have similar decay rates ). The hypergeometric function scales as order $n_f^{1-p_f}$, and it can be shown that this leads to $I_R(p_f+1)/I_R(p_f) \sim n_f/(n_f+1)$. Thus for $n_f \sim 10^{34}$, $I_R$ ( $p_f=1$, $\Delta p=0$, $\Delta n=1$) differs from $I_R$ for other $p=p_f \neq 1$ diagonals by less than 1 part in $10^{34}$ per unit $p$-diagonal change provided $p << n$. Hence corresponding transition rates for constant-$p$ dipole transitions for deeper states such as E to F of figure 1 are essentially the same as for an A to B transition, provided $p << n_f$. It remains therefore to examine the transition rates of the other decay channels contributing towards the net decay rate of any given low-$p$ state. These involve transitions where $\Delta n = 2, 3, 4 \ldots$.

### 4.3. Transitions where $\Delta n = 2, 3, 4 \ldots$

We take equation (15) and modify it for large $n_f$ using Stirling's formula to avoid terms in $n_f!$:

$$I_R = \frac{2^{n_f - p_f + \frac{5}{4}} n_f \left(n_f^2 + n_f \Delta n\right)^{n_f - p_f + 2} \left(2n_f - p_f + 1\right)^{n_f - p_f/2 + 1/4} b_0}{\exp\left((p_f-3)/2\right) \left(2n_f + \Delta n\right)^{2n_f - 2p_f + 5} \left(n_f - p_f + 2\right)^{n_f - p_f + 7/4}} \times$$

$$\left(\frac{\left(p_f - 1\right)!}{\left(p_f + \Delta n - 2\right)!}\right)^{\frac{1}{2}} \times \left(\prod_{i=1}^{\Delta n + p_f - 2} \left(2n_f - 2p_f + i + 3\right)\right)^{\frac{1}{2}} \times \quad (17)$$

$$\sum_{k_f=0}^{p_f-1} \left(\frac{\dfrac{(2n_f - 2p_f + k_f + 2)(2n_f - 2p_f + k_f + 3)(2n_f - 2p_f + k_f + 4)}{k_f!(p_f - k_f - 1)!}}{\dfrac{(-2n_f - 2\Delta n)^{k_f}}{(2n_f + \Delta n)^{k_f}} \, {}_2F_1\left(a, b; c; z\right)}\right)$$

This equation includes terms of order $n_f^{n_f}$ which cannot be calculated when $n_f$ is large. Provided however that $n_f >> \Delta n$ and $n_f >> p_f$, equation (17) may be put into a suitable approximate form, using the fact that

$$\lim_{n_f \to \infty} \left(\left(1 + \frac{\Delta n}{n_f}\right)\left(1 + \frac{(1-p_f)}{2n_f}\right)\left(1 + \frac{\Delta n}{2n_f}\right)^{-2}\left(1 + \frac{(2-p_f)}{n_f}\right)^{-1}\right)^{n_f} = e^{\frac{1}{2}(p_f - 3)}$$

which gives

$$I_R = \frac{\left(2n_f\right)^{3-p_f}\left(n_f+\Delta n\right)^{2-p_f} b_0}{2^{7/4}\left(2n_f+\Delta n\right)^{5-2p_f}} \times$$

$$\left(\frac{\left(p_f-1\right)!\left(2n_f-p_f+1\right)^{\frac{1}{2}-p_f}}{\left(p_f+\Delta n-2\right)!\left(n_f-p_f+2\right)^{\frac{7}{2}-2p_f}}\prod_{i=1}^{\Delta n+p_f-2}\left(2n_f-2p_f+i+3\right)\right)^{\frac{1}{2}} \times \qquad (18)$$

$$\sum_{k_f=0}^{p_f-1}\left(\frac{\dfrac{(2n_f-2p_f+k_f+2)(2n_f-2p_f+k_f+3)(2n_f-2p_f+k_f+4)}{k_f!(p_f-k_f-1)!}}{\dfrac{(-2n_f-2\Delta n)^{k_f}}{(2n_f+\Delta n)^{k_f}}\,{}_2F_1\left(a,b;c;z\right)}\right)$$

or equivalent alternate expressions involving various combinations of the other related parameters such as $n_i$, $p_i$ etc.

Equations like (18) not only enable easier approximate calculations of $\Delta n = 1$, $\Delta p = 0$ transitions than does equation (16), but also transitions where $\Delta n > 1$ and ones which do not necessarily end on the $p_f = 1$ diagonal, provided that $p_f$ and $\Delta n$ remain small compared to $n_f$. (The upper limit on $p_f$ for $I_R$ being calculable is determined by the number of terms one can tolerate in the summation and number of significant figures that are needed to avoid summation truncation errors.) The accuracy of the approximation can be estimated by the error associated with the above limit and that for Stirling's approximation, which were used in its derivation. (For $n_f \sim 10^4$ the error in equation (18) is less than 1 part in 3000 when $\Delta n$ and $p_f < 10$.)

### 4.4. Decay rates for states where $p \neq 1$

For the deeper ($p > 1$) states of figure 1 with many decay channels, we designate the $\Delta n = 1$ transition as the 'primary' decay channel. It might be expected that deep states could have so many channels available that the net decay rate could be much higher than that due to the primary $\Delta n = 1$ transition alone, especially given that these channels correspond to larger values of $\Delta n$ which result in proportionally larger values of $\omega$ in a transition rate formula which involves a cubic dependence on $\omega$. For a given deep state of fixed level $n = n_i$ lying on a diagonal $p = p_i$ we wish to be able to calculate transition rates for each transition of the ensemble of $\Delta l = 1$ channels available, starting with $\Delta n = 1$:

$(n_i, p_i) \rightarrow$ $(n_f = n_i - 1, p_f = p_i)$ and continuing to $\Delta n = p_i$: $(n_i, p_i) \rightarrow$ $(n_f = n_i - p_i, p_f = 1)$ (and an equivalent set for $\Delta l = -1$). For $n_i$ and $n_f$ values that might be significant in an astrophysical context (i.e. $\geq 10^{25}$), $I_R$ obtained using equation (18) is realistically calculable for $p$ values up to several thousand). Figure 6 shows decay rates of all the allowed ($\Delta l = 1$) channels for each of 9 initial states ($n_i = 10^{30}, p_i = 25, 50, 100$), ($n_i = 10^{34}, p_i = 25, 50, 100$) and ($n_i = 10^{34}, p_i = 25, 50, 100$). The dramatic decrease in transition rate $1/\tau$ demonstrates that the $\Delta n = 1$ is indeed the predominant decay channel of any high-*n* low-*p* state.

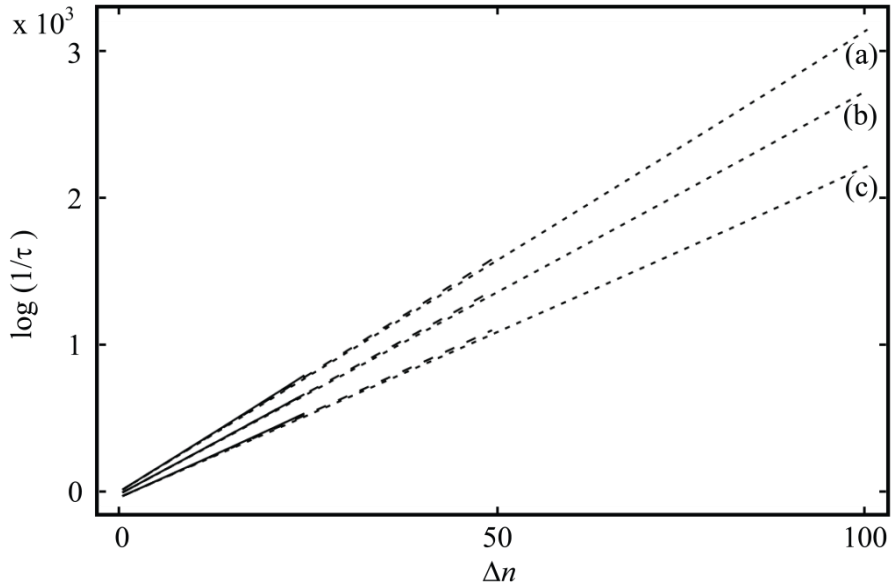

**Figure 6.** Decay rates $(1/\tau)$ of all available channels of the initial low-*p* states $p_i = 100$ (short dashes), $p_i = 50$ (long dashes) and $p_i = 25$ (solid curve) for an initial quantum level (a) $n_i = 1.0$ x $10^{34}$; (b) $n_i = 1.0$ x $10^{30}$; and (c) $n_i$ = 1.0 x $10^{25}$; plotted as log$(1/\tau)$ versus the quantum jump $\Delta n$;
Parameter values: central mass = 2.0 x $10^{42}$ kg; particle mass = 1.67 x $10^{-27}$ kg; charge = 1.6 x $10^{-19}$ C;

The decay rate of any particular channel is determined predominantly by the value of $I_R$ as calculated by equation (18). Its use is limited to low $p$ values, chiefly because of the number of significant figures required to avoid truncation errors, and the fact that the summation component requires $\sim p$ terms. However provided that $n_i \gg 1$, estimates of channel decay rates may also be determined by considering an equivalent transition having the same $\Delta p$ but one that ends on $p_f = 1$ (*c.f.* equation (16) above). The value of $I_R$ for these $p_f = 1$ transitions may be obtained from $I_R = b_0 n_i \left( \dfrac{2}{\pi p_i} \right)^{\frac{1}{4}} \left( \dfrac{e}{2} \right)^{\frac{p_i}{2}} \left( \dfrac{p_i}{n_i} \right)^{\frac{p_i - 3}{2}}$, valid when

$n_i \gg p_i \gg 1$, the logarithm of which is calculable for all $p_i$ and $n_i$.

Figure 7 shows the variation of $\log(I_R)$ as a function of $\Delta p = p_i - 1 = (\Delta n - 1)$ for dipole transitions ending on $(n_f, p_f = 1)$ with $n_f$ remaining constant, as a function of the difference in the $p$ value between the initial and final states. As expected, the values of $I_R$ for low $\Delta n$ are similar to their counterparts that were used to calculate the transition decay rates shown in figure 6. For the large values of $n_i$ (or $n_f$) considered here it can be shown from equation (16) that the ratio $I_R(p=2)/I_R(p=1)$ is $\sqrt{2n_i}$. A similar trend continues: while $p \ll n$ large, the ratio $I_R(p+1)/I_R(p)$, is given approximately by (using equation (18)) $I_R \sim \left( \sqrt{p/2n_i} \right) \left( (p+1)/p \right)^{p-1} \sim b_0 \sqrt{p/2n_i}$ and demonstrates that $I_R$ decreases dramatically (because of the large values of $n_i$ and $n_f$) for each successive unit increase in $p$. This results in the successive values for the overlap integral $\Pi_{if}$ being also reduced by successive factors of $\sim \sqrt{p/2n_i}$ for successive channels originating from the fixed upper state (because provided $p$ is small, we can relate each decay channel from the given initial state to an equivalent, similarly-lifetimed transition that ends on a $p = 1$).

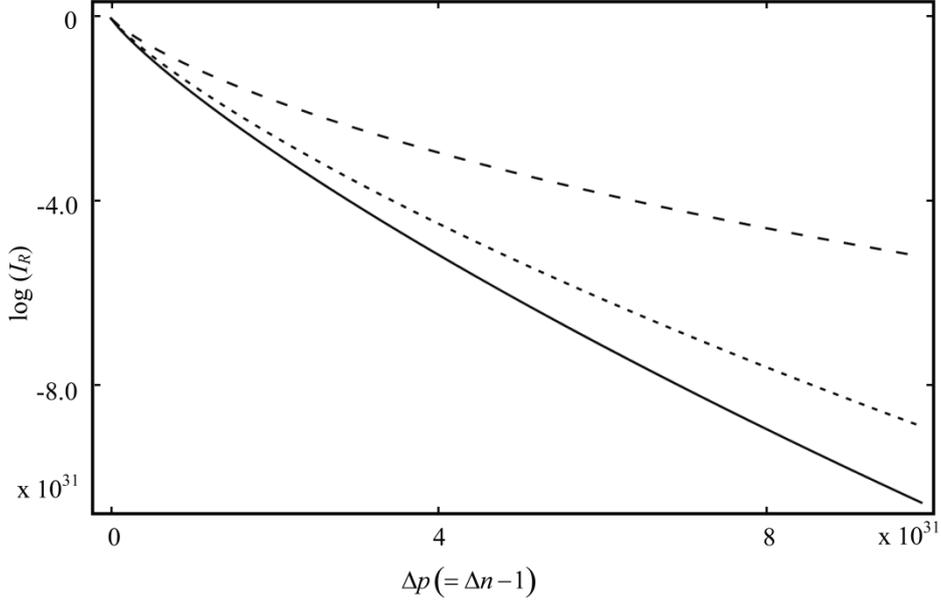

**Figure 7.** Variation of the radial component of the dipole matrix element plotted as a function of $\log(I_R)$ versus the difference in the $p$ value between an initial ( $p_i$ ) and final state ( $p_f = 1$ ).: $n_f = 10^{34}$ (solid curve); $n_f = 5 \times 10^{33}$ (short dashes); $n_f = 10^{33}$ (long dashes).
Parameter values: central mass = 2.0 x $10^{42}$ kg; particle mass = 1.67 x $10^{-27}$ kg; charge = 1.6 x $10^{-19}$ C;

The trend $I_R \sim \left( \sqrt{p / 2n_i} \right) \left( (p+1) / p \right)^{p-1}$ does not continue once $p$ becomes a substantial fraction of $n_f$. Using equation (18) a better but more complex approximation than $I_R = b_0 n_i \left( \dfrac{2}{\pi p_i} \right)^{\frac{1}{4}} \left( \dfrac{e}{2} \right)^{\frac{p_i}{2}} \left( \dfrac{p_i}{n_i} \right)^{\frac{p_i - 3}{2}}$ for transitions ending on the $p_f = 1$, and without the restriction $n_i \gg p_i \gg 1$ (although $n_i \gg 1$ still a requirement) can be obtained as

$$I_R = \frac{2^{n_f 2} b_0 \left( p_i + n_f \right)^{n_f + 2}}{\sqrt{e}} \left( \frac{n_f^{\,9}}{\pi \, p_i^{\,7-2j}} \right)^{\frac{1}{4}} \frac{\left( 2n_f + p_i + 1 \right)^{\frac{4n_f + 2p_i + 1}{4}}}{\left( 2n_f + p_i \right)^{2n_f + p_i + 2}} \qquad (19)$$

where $n_f = n_i + p$, and from this equation it may shown that $I_R(p+1)/I_R(p) \to 1$ as $p \to \infty$ so that the $\ln(I_R)$ vs $p$ plot of this function approaches a horizontal asymptote.

Nevertheless, provided $p_i (= \Delta n - 1)$ remains small compared to $n_i$, the above discussion implies that the state-to-state transition rate $(1/\tau)$ for successively lower channels (i.e. as $\Delta n$ increases) is $\propto \omega^3 \Pi_{if}^2 \propto \Delta n^3 (\Delta n / 2n_i)^{\Delta n}$ and also decreases rapidly. For example in the case of $n_i = 10^{34}$, for an upper, moderately deep state (i.e. one whose initial state $p$ value $p_i$ is still much less than its $n$ value $n_i$) the value of $I_R$ (and $\Pi_{if}$) for the $\Delta p = 1$ transition channel will be $\sim 10^{-17}$ times that for the corresponding $\Delta p = 0$ transition channel, but $\omega^3$ will increase only by a factor of 8, and the transition rate for the $\Delta p = 1$ transition will therefore be $\sim 10^{33}$ times smaller than for the $\Delta p = 0$ transition.

The net transition rate for such a deep state $(n_i, l_i = n_i - p)$ with $n_i \gg p \gg 1$ is then the sum of all the individual state-to-state transition rates, $\sum_{\Delta n = 1}^{p} A_{(n_i, n_i - p),(n_i - \Delta n, n_i - p - 1)}$ where $A_{(n_i, n_i - p),(n_i - \Delta n, n_i - p - 1)}$ is the state-to-state transition rate for the each of the respective decay channels. When $n_i$ is large the only significant term in this summation is the primary diagonal one $\Delta n = 1, \Delta p = 0$. Thus provided $p \ll n_i$ and $n_i \gg 1$, the lifetimes of any state whose parameter values are $(n_i, l_i = n_i - p)$ where $n_i \gg p \geq 1$, are approximately the same as the single state-to-state transition time of the primary diagonal channel for that state decay, that is the $\Delta n = 1, \Delta p = 0$ transition. This then is itself very similar in lifetime to that of the equivalent $\Delta n = 1, \Delta p = 0$, $p = 1$ state given in Table 2.

If the condition on $p \ll n_i$ is not satisfied transition decay rates increase dramatically with increasing $\Delta n = 1$ because of the factor $\omega^3$, and there is opportunity for a states to rapidly decay given the number of channels available. For example from figure 8 it can be seen that the value of $I_R$ (and hence the decay rate) for a given constant $\Delta n = n_i - n_f$ increases to a non-trivial value once the initial $p$ value is a significant fraction of $n$. The effect is strongly dependent on $n_i$ (or equivalently $n_f$) and the larger $n_i$ the greater the relative increase in $I_R$ on moving horizontally across the *n-p* state diagram.

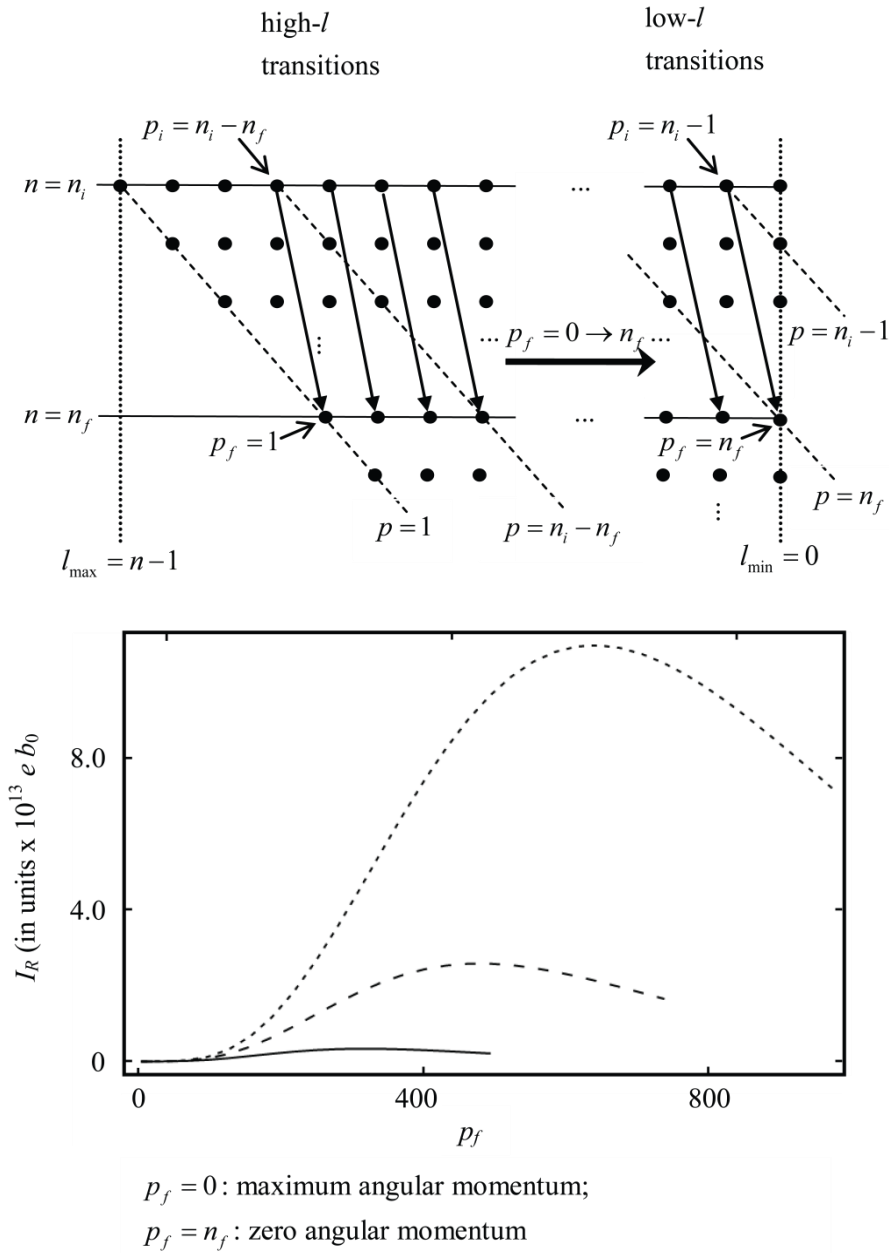

$p_f = 0$: maximum angular momentum;

$p_f = n_f$: zero angular momentum

**Figure 8.** Variation of the radial dipole integral $I_R$ with the horizontal position of the transition on the $n$-$p$ state diagram. $\Delta n$, $\Delta l$ and $n_f$ are constant for each curve. small dashes: $n_i = 1000$, $n_f = 990$; large dashes: $n_i = 750$, $n_f = 740$; solid: $n_i = 500$, $n_f = 490$.

The trend displayed here, that $I_R$ has a significantly high value when $p$ is a substantial fraction of $n$, means that decay rates of these low angular momentum states have potential to be very large. In contrast to the states where $p$ is a negligible fraction of $n$ and multichannel decay is insignificant, this result demonstrates the critical importance of multichannel decay in low-$l$ cases (and also in low-$n_i$ cases). Furthermore it enables us to understand the differences in interaction properties of the specialised, low-interaction, high-$n$, low-$p$ states and those of the highly mixed states of more localised particles that have extremely broad eigenspectra, and how the Compton cross-section, for example, can be different for a set of such low-$p$, high-$n$ eigenstates compared to that obtained from the Klein-Nishina equation, from an ionised gas cloud or measured in the lab.

### 4.5. Decay rates for states where $p \gg 1$

When both initial and final states have $p \gg 1$ then no approximations discussed so far can be used to estimate $I_R$. It may still be possible however to gain information about the value of $I_R$ using knowledge of the position, extent and spatial oscillation frequencies of the state functions, found by solving equation (14). This could be particularly useful in particle interactions where one is concerned to know the range of states into which any given eigenstate might be scattered by interactions with a traversing particle, or the degree to which that traversing particle might itself be scattered.

There is an interesting empirical relationship between the value of $I_R$ and the relative spatial oscillation frequencies (RSOF) of the two states involved in the overlap. Figure 9 shows that for transitions where one of the states has $p = 1$ the value of $\log\left(-\log\left(I_R\right)\right)$ is linearly related to the log(RSOF) of the two states. This is a direct consequence of similar relationships that exist between $p$ and $I_R$, and $p$ and the spatial extent and spatial oscillation frequency of the wavefunction shown in the insets. Thus given the $p$ values of two states and the types of relationships shown in figure 9, it should be possible to predict their relative extents and relative spatial oscillation frequencies and from these obtain estimates of $I_R$ for these states when they overlap. The significant thing about this approach to estimating values of $I_R$ is that it has a much wider scope. Although it is a much more approximate method of determining upper limit estimates of overlap integrals, it has the advantage of depending only on the properties of the states themselves and not on the type of interaction potential. Thus it may be used as "first guess" or estimate to determine whether a decay or transition rate will be insignificant for other types of interactions as well, for example non-dipole decays and particle interactions. If the spatial extents and oscillation frequencies of two states differ by large amounts then the probability of any interaction inducing transitions from one of these states to the other will be small even if the states overlap in space.

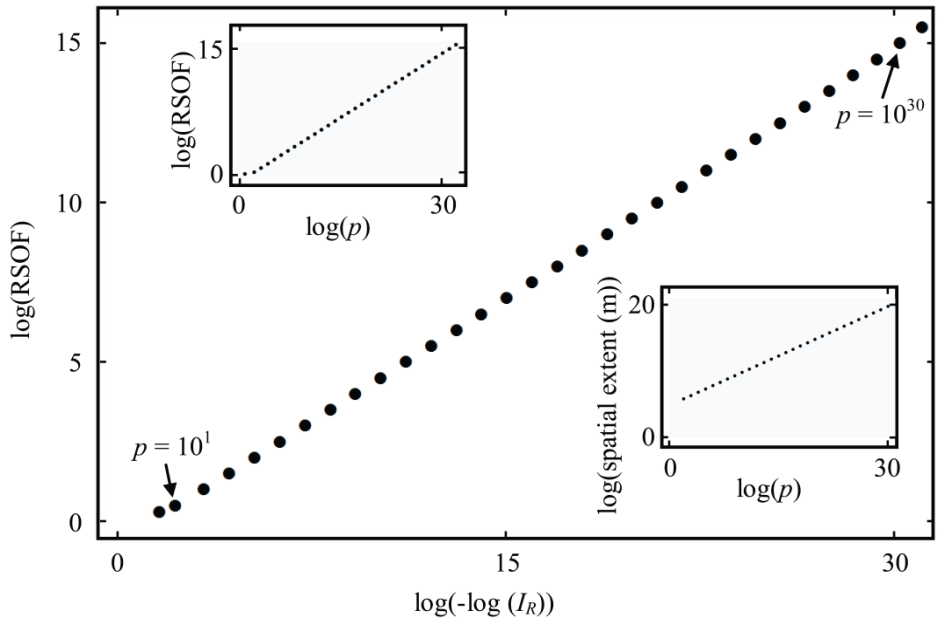

**Figure 9.** The value of $\log\!\left(-\log\!\left(I_R\right)\right)$ versus $\log(\text{RSOF})$ for transitions where one state has $p=1$ and the other $p=4,\ 10,\ 10^2,\ 10^3,$ ... $10^{31}$ (see text). Insets show the variation of the radial spatial extent (radial width) and RSOF with $p$. Parameter values: central mass = 2.0 x $10^{42}$ kg; particle mass = 1.67 x $10^{-27}$ kg;

### 4.6. Invisibility and the weakly interacting nature of the low-p eigenstates

The extreme longevity of the low-$p$, high-$n$ eigenstates when in large central mass potentials has consequences beyond that of negligible electromagnetic radiation emission, for example their inability to effectively undergo gravitational collapse. More importantly however they will be weakly interacting with respect to a large portion of the electromagnetic spectrum. They will therefore be essentially invisible as may be shown from the following examples.

We consider an incident flux of electromagnetic radiation impinging on an assembly of eigenstates and ask firstly, what transition rates does this radiation produce among the eigenstates, and secondly, what effects do the eigenstates produce on the traversing beam of radiation? The relevant processes for bound eigenstates are those of stimulated emission and absorption. (Classical electromagnetic scattering processes from 'free-particles' (Raleigh, Compton, etc.) may also be analysed using the summation of interactions over a vast eigenstate ensemble but the significant interaction channels for these processes do not involve channels that incorporate the low-$p$ eigenstates.) The probability per unit time $P_{if}$ of stimulated emission or absorption of radiation from a state $i$ to a state $f$ is given by [5]

$$P_{if} = \frac{\pi e^2}{3\varepsilon_0 \hbar^2} \left| \left\langle f \left| \mathbf{r} \right| i \right\rangle \right|^2 \rho\left(\omega_{if}\right) = \frac{\pi \Pi_{if}^{\ 2}}{3\varepsilon_0 \hbar^2} \rho\left(\omega_{if}\right) = \frac{4\pi^2 \Pi_{if}^{\ 2}}{3\varepsilon_0 \hbar^2 c} I\left(\omega_{if}\right) \qquad (20)$$

where $\rho(\omega_{if})$ is the radiation energy density per unit angular frequency and $I(\omega_{if})$ is the corresponding beam intensity per unit angular frequency. Technically, for degenerate states, (20) must be summed over the available final states and averaged over the initial ones but since the significant transitions are dipole, the relevant selection rules mean that we may consider the effective degeneracies of the initial and final levels to be approximately the same and take (20) to be a reasonable approximation to the net eigenstate transition rate in the following examples.

As a first example, we consider photon absorption/stimulated emission from high-$n$, $p=1$ eigenstates. $\Pi_{if}$ is largest for $\Delta n = \pm 1$ (see equation (27) of [1]). The most pervasive radiation extending over the entire volume of the eigenstates is that of the CMB and at the resonant frequency at present day CMB densities, $\rho\left(\omega_{if}\right) \sim 7 \times 10^{-82} \, \mathrm{J\,m^{-3}\,s\,rad^{-1}}$. Taking a typical proton-occupied eigenstate with $n \sim 5 \times 10^{33}$ in a gravitational potential of central mass $\sim 10^{42}$ kg gives $P_{if} \sim 4 \times 10^2 \, \mathrm{s^{-1}}$. We thus expect that any typical $p=1$ eigenstate to be subjected to random promotion/demotion state fluctuations of this magnitude. Over a cosmic time of $10^{17}$ s, any given state might be expected to have dispersed over $\sqrt{4 \times 10^{19}} \sim 10^{10}$ $n$-levels, clearly negligible compared to the value of $n$ itself. Even allowing for an increased intensity of $\rho(\omega_{if})$ over cosmic history by using $\rho\left(\omega_{if}\right) \sim 7 \times 10^{-78} \, \mathrm{J\,m^{-3}\,s\,rad^{-1}}$ (at $z=10000$) gives a $n$-level dispersion of $\sim 10^{12}$ over the cosmic history. Similar values will be obtained for other low-$p$, $\Delta p = 0$, transitions.

There are of course many promotion channels ($\Delta n > 1$) available for excitation of $p=1$ (or other low-$p$) eigenstates. However, as demonstrated by equation

(27) of [1] and equation (18) of this present paper, the value of the radial part of $\Pi_{if}$ decreases so rapidly with increasing $\Delta n$, that the contribution to the excitation of a low-$p$ eigenstate by stimulated absorption is dominated by the $\Delta p = 0$ channel.

The second example involves calculation of the numbers of photons that would be absorbed or scattered in crossing a typical halo. We consider a beam of photons of number density $n_\gamma$ m$^{-3}$ whose frequencies are distributed uniformly within a small angular frequency range $\Delta \omega$. The beam intensity per unit angular frequency $I(\omega)$ is $I(\omega) = n_\gamma c \hbar \omega / \Delta \omega$, enabling (20) to be rewritten as

$$P_{if} = \frac{4\pi^2 \, \Pi_{if}{}^2 n_\gamma \omega}{3\varepsilon_0 \hbar \, \Delta \omega} \tag{21}$$

where $P_{if}$ represents the probability per unit time for the promotion of a single eigenstate. We assume that $n_\gamma$ is uniform throughout a typical halo. The number of excitations per second within the halo is then given by $P_{if}$ times the number of available eigenstates $N_s$ that have transitions whose frequencies lie within the angular spread $\Delta \omega$. For the purposes of illustration we assume that the halo consists of an eigenstructure filled from the longest lived $p = 1$ states to some maximum $p$ level $p_{\max}$. To accommodate a halo mass of $\sim 10^{42}$ kg with radius $\sim 1.5 \times 10^{21}$ m requires totally filling all levels from $p = 1$ to $p = 10$.

We now consider the transition $\Delta n = 1, \Delta p = 0$, originating from the state $n = 5 \times 10^{33}$, $p = 1$. For a given value of $n$ and $p$, there are $2l + 1 \sim 10^{34}$ z-projection substates available for promotion with the same central frequency. It may also be shown that $\Delta p = 0$ transitions originating on the states $n = 5 \times 10^{33}$, $p = 2, \ldots 10$ have profiles that lie essentially within the natural linewidth of the $n = 5 \times 10^{33}$, $p = 1$, $\Delta n = 1, \Delta p = 0$ transition, and that so do states whose $n$-values lie within $\sim 1 \times 10^{24}$ of $n = 5 \times 10^{33}$. This means that if we take the spread $\Delta \omega$ of the $n_\gamma$ m$^{-3}$ photons to be within the natural linewidth of the transition, there will be $10^{34} \times 10 \times 10^{24} = 10^{59}$ states available for promotion within the angular frequency spread of the photon beam, so that $N_s \sim 10^{59}$. Since every time a transition is excited by absorption a photon is removed from the field we may then, using (21) write down a rate $dN_\gamma / dt$ at which photons are removed from the photon beam across the whole halo as

$$\frac{dN_\gamma}{dt} = -\frac{4\pi^2 \Pi_{if}^2 n_\gamma \omega N_s}{3\varepsilon_0 h \Delta \omega} \tag{22}$$

or by writing $N_\gamma = n_\gamma V$ where $V$ is the halo volume, obtain the expression

$dn_\gamma = -(1/\tau) n_\gamma dt$, where $\tau$ is the expected lifetime for a photon of frequency

$\omega$ to survive without absorption, given by

$$\tau = \frac{3\varepsilon_0 h V \Delta \omega}{4\pi^2 \Pi_{if}^2 \omega N_s} \tag{23}$$

For a transition $\Delta n = 1, \Delta p = 0$, originating from $n = 5 \times 10^{33}$, $p = 1$, $\tau \sim 10^{-56}$ s so that photons of frequency $\sim 10^{-16}$ Hz would be readily absorbed as would be expected from the first example. For $\Delta n = 2, \Delta p = 1$, originating from $n = 5 \times 10^{33}$, $p = 1$, the absorption frequency is twice as great but $\Pi_{if}$ is much reduced so that $\tau \sim 10^{-22}$ s. For $\Delta n = 3, \Delta p = 2$ the lifetime is of the order of the galactic halo transit time ($\tau \sim 10^{12}$ s) and for $\Delta n \geq 4, \Delta p = 2$ the photon lifetimes become much greater than the halo transit time. Thus for any photon frequency realistically observable within the electromagnetic spectrum ($\omega \geq 1\,\text{kHz}$ implies $\Delta n \geq 10^{18}$), $p = 1$ eigenstates do not absorb radiation. Provided the $p$-value of the initial state is small, and certainly for $p < 10^9$, similar results would apply, leading to the invisibility of all but the highest $p$-valued sates. Therefore provided that an eigenstructure consisted predominantly of particles occupying low-$p$ eigenstates, it will be transparent to virtually all regions of the electromagnetic spectrum including the optical region.

The weak response of the low-$p$ eigenstates to stimulated emission and absorption comes about directly as a consequence of the rapid decrease in the value of $I_R$ with increasing RSOF between initial and final eigenstates, discussed in the previous section. Similar cancellation effects within the overlap integrals can be seen in the angular domain as well and whenever $\Delta p$, $\Delta n$, $\Delta m$ or $\Delta l \gg 1$ there can be significant differences in spatial oscillation frequencies between initial and final states resulting in weak interaction often irrespective of the Hamiltonian involved. Given an initial high-$n$, low-$p$ state on figure 1, one can in general one can make a distinction between several types of transitions depending on the location of the final state on figure 1. In the case of photon absorption discussed above for example, the final state lies almost vertically above the initial state (since $\Delta l = \pm 1$) and there is always a large change in p giving rise to large RSOFs and infinitesimal values for $I_R$. We can also consider interactions giving rise to large horizontal state changes. This corresponds to elastic scattering: large changes to the momentum of both the eigenstate and the perturbing entity occur without energy transfer. Necessarily large changes in $p$ are involved with correspondingly low overlap integrals and hence limited opportunity for interaction unless the selection rules are such as to permit a large range of available channels. It is clear that when processes such as elastic and Compton scattering are described in the present basis set of eigenfunctions, then these processes must be seen to take place largely through mixtures of the low-$l$, high-$p$ channels that enforce the traditional localization the particles and give them opportunity to interact in the classical sense.

To avoid changes in $p$ and ensure non-negligible overlap integrals, interactions need to involve initial and final states that lie on relatively closely spaced diagonals. For low-$p$ states this presents a problem for interaction. The fact that the radial eigenfunctions decrease rapidly outside their range, coupled with their relatively small radial spreads and positions that vary significantly with $n$, enforces a limit on the extent to which momentum and energy can be transferred. If the initial and final states differ too much in energy and/or momentum then they will cease to overlap and interaction integral will be zero, irrespective of the potential involved. This is major reason for the weakly interacting nature of the low-p states.

Equation (14) is useful in determining when two given states will cease to overlap and in defining the menaing of 'low-$p$'. For example given two deeper-$p$ states (but still with the condition that $p << n$) we may ask a similar question that posed earlier for transitions between two $p = 1$ states, that is: how far away must such states be on the $n$-$p$ diagram to ensure that 'no' transitions occur between them irrespective of the interaction potential involved? Essentially this involves comparing the spatial position of the lowest zero of the radial function of the upper level with the spatial position of the largest zero of the radial function of the lower level. If there is a substantial positive difference between these two values then we can be reasonably confident that the radial functions do not overlap because the mathematical behaviour of the radial eigenfunction is to decrease rapidly for values outside its radial range. This work is still in progress and will be reported on at a later date. We do expect however that cross sections for interactions between particles and eigenstates will be larger than those with photons, because of the less restrictive selection rules for interactions involving particles. (That is low-$p$ eigenstates may not be as "invisible" for particles as they are for photons.)

We emphasise one final point about the measurement of cross sections. The classical elastic or inelastic photon scattering processes (for example Compton and Rayleigh scattering) have measured cross sections determined from experiments carried out using particle assemblies in vessels that contain macroscopic numbers of particles localized within the vessel. These 'localized' particles necessarily have eigenspectra that, by virtue of this localization, are composed of a vast but reproducible summation of the much more highly interacting low-$l$, high-$p$ states, the makeup of which is determined by macroscopic parameters such as temperature, number density etc. Individual eigenstates or specialised arrays of eigenstates with unique features will have the capacity to exhibit significantly different properties from say, a gas in thermal equilibrium (in the same way for example that the scattering cross section for a atomically bound electron differs from that of a free electron). Any measurement of a particular cross section should necessarily be accompanied by reference to the relevant time-averaged eigenspectra of the particle ensemble under which the measurement was made. This is done automatically for example when we recognise the dependency of cross sectional data on external parameters such as temperature etc. that determine the eigenspectral mix.

## 5. Conclusion

This paper has developed further, more approximate procedures for studying the high-$n$, high-$l$, low-$p$ gravitational eigenstates. This has included investigation of the trends in functional behaviour of both the angular and radial components of high quantum number eigenfunctions and development of a formula for calculating the width, position and spatial oscillation frequencies of the zeros of the radial component. These trends have enabled many potential dipole decay channels to be eliminated as effective channels and hence facilitated estimation of the net dipole decay rates of many 'deep', high-$n$ states even when those states may have many channels available for decay.

The first point of interest is the fact that the spatial oscillation frequencies of the eigenfunctions increase in proportion to the square root of the $p$-value, which is the dominant physical explanation as to why the radial overlap integral for high-$p$ with low-$p$ states decreases so rapidly with increasing $p$, irrespective of the interaction potential and the degree of state overlap. It means that not only do the low-$p$ eigenstates have extremely long decay times and gravitational stability, but also that they cannot easily be pushed into higher levels by stimulated transitions or (to a lesser extent) transferred to other locations with very different $p$ values on the state diagram by other interactions such as particle collisions. This accounts for their invisibility and their generally weakly interacting nature.

An important consequence of the dramatic effects of spatial oscillation frequencies is that high-$n$ transitions must necessarily take place along diagonal lines of approximately constant-$p$. For the high-$n$, low-$p$ states, the only transitions possible are those whereby they are transferred to other low-$p$ states. For these types of transitions however, the role of the distinct width, position and rapid cut-offs of the states becomes important. This restricts the amount of energy ($n$) and momentum ($l$) that can be exchanged in any interaction involving a low-$p$ state and consequently also restricts the changes that take place in the perturbing entity and again contributes to the eigenstate's weakly interacting behaviour. On the other hand the low-$l$ states have large radial extents and interactions that approximately conserve $p$ can take place with a much greater exchange of energy and momentum. This explains how the cross-sections for traditional processes like elastic and Compton scattering can be quite different for the low-$p$ eigenstates.

It is also interesting to note that the weakly interacting properties of the particles occupying the low-$p$ eigenstates arise as a consequence of the quantum states in which they are placed and not from any internal weakly interacting properties of the particles themselves, a fact that is highlighted by the present work. Indeed because the weak interaction is a consequence of the properties of the wavefunctions themselves and their behaviour on the scales and sizes of the quantum parameters involved, the phenomena is not restricted to gravitational potentials and general macroscopic low-$p$ eigenfunctions could in some sense provide us with an unusual structural form of matter that we are yet to become familiar with.

The prediction of invisibility and weak interaction of the specialised gravitational eigenstates is undoubtedly the most significant result to come out of the present work. If such gravitational eigenstates exist then they may be responsible for most of the dark matter in the universe, especially if, as it appears, a viable primordial-nucleosynthesis-consistent scenario for their production might be possible [2]. This would make them particularly attractive as an explanation for dark matter, since it means that its existence could be explained without the need to introduce any new particles and that its origin follows as an automatic prediction from traditional quantum theory.